\documentclass[natbib,printer]{aa}
\usepackage[varg]{txfonts}
\usepackage{natbib}
\usepackage[switch,modulo]{lineno}
\bibpunct{(}{)}{;}{a}{}{,} 
\usepackage{color}
\definecolor{todo}{rgb}{0.8,0.2,0.3}

\definecolor{comm}{rgb}{0,0.7,0}

\definecolor{idea}{rgb}{0,0.5,1}

\usepackage{longtable}

\begin{document}
\title{HD\,133729: A blue large-amplitude pulsator in orbit around a main-sequence B-type star}
\titlerunning{BLAP in orbit around a main-sequence star}
\author{A.~Pigulski\and K.~Kotysz \and P.~A.~Ko{\l}aczek-Szyma\'nski
}
\institute{Instytut Astronomiczny, Wydzia{\l} Fizyki i Astronomii, Uniwersytet Wroc{\l}awski, Kopernika 11, 51-622 Wroc{\l}aw, Poland \\
\email{Andrzej.Pigulski@uwr.edu.pl}
\label{inst1}
}
\date{Received date / Accepted date }
\abstract{
Blue large-amplitude pulsators (BLAPs) form a small group of hot objects pulsating in a fundamental radial mode with periods of the order of 30 minutes. Proposed evolutionary scenarios explain them as evolved low-mass stars: either $\sim$0.3\,M$_\odot$ shell-hydrogen-burning objects with a degenerated helium core, or more massive (0.5\,--\,0.8)\,M$_\odot$ core-helium-burning stars, or $\sim$0.7\,M$_\odot$ surviving companions of type Ia supernovae. Therefore, their origin remains to be established. Using data from Transiting Exoplanet Survey Satellite, we discovered that HD\,133729 is a binary consisting of a late B-type main-sequence star and a BLAP. The BLAP pulsates with a period of 32.37\,min decreasing at a rate of $(-7.11 \pm 0.33)\times 10^{-11}$. The light curve is typical for BLAPs, but shows an unusual 40-s long drop at the descending branch. Due to light dilution by a brighter companion, the observed amplitude of pulsation is much smaller than in other BLAPs. From available photometry, we derived times of maximum light, which revealed the binary nature of the star via $O-C$ diagram. The diagram shows variations with a period of 23.08433~d that we attribute to the light-travel-time effect in the system. The analysis of these variations allowed to derive the spectroscopic parameters of the BLAP's orbit around the center of the mass of the binary. The presence of a hot companion in the system was confirmed by the analysis of its spectral energy distribution, which was also used to place the components in the Hertzsprung-Russell diagram. The obtained position of the BLAP fully agrees with the location of the other members of the class. With the estimated $V\approx 11$\,mag and the Gaia distance of less than 0.5\,kpc, the BLAP is the brightest and the nearest of all known BLAPs. It may become a clue object in the verification of the evolutionary scenarios for this class of variable. We argue that low-mass progenitors of the BLAP are excluded if the components are coeval and no mass transfer between the components took place.
}
\maketitle
\authorrunning{A.~Pigulski et al.}
\section{Introduction}
Large photometric ground-based surveys provided astronomical community with a huge number of photometric measurements of millions of stars. Space missions were another source of a large quantity of photometric data, some with an unprecedented quality and time coverage. Although the surveys and the missions were primarily aimed at a detection of microlensing events and transiting exoplanets, the data occurred to be extremely useful in the characterisation of the different types of pulsating stars over the entire Hertzsprung-Russell (H-R) diagram. No wonder some exotic and very rare types of variable stars were found. An example is the discovery of a small group of short-period hot pulsators known as blue large-amplitude pulsators, hereafter BLAPs \citep{2017NatAs...1E.166P}.

The first BLAP was discovered serendipitously by \cite{2013AcA....63..379P} in the Optical Gravitational Lensing Experiment (OGLE) data while searching for pulsating stars in the OGLE-III Galactic disk fields in the direction tangent to the Centaurus Arm. The star, OGLE-GD-DSCT-0058, showed large peak-to-peak amplitude ($\Delta I=0.24$\,mag) and light curve resembling fundamental-mode pulsators related to $\delta$~Scuti stars, that is, high-amplitude $\delta$~Scuti (HADS) stars and SX Phoenicis stars. The star was therefore tentatively classified as a $\delta$~Scuti variable. It differed from the members of these groups, however, by its exceptionally short period (28.26\,min). The follow-up spectroscopic study \citep{2015AcA....65...63P} showed that it cannot be a $\delta$~Scuti star because it is too hot -- its effective temperature amounted to about 33\,000~K. Searching for the other members of this group in the Galactic bulge OGLE data, \cite{2017NatAs...1E.166P} found 13 BLAPs. This allowed the characterisation of this new class of pulsating stars. BLAPs pulsate with periods between 22 and 40 minutes and have non-sinusoidal (a steep increase followed by a slow decrease of brightness) light curves with the peak-to-peak amplitudes of 0.2\,--\,0.4\,mag in the $V$ band. Amplitudes in the $I$ band are on average about 20\% lower than in the $V$ band, which is typical for radially pulsating stars. In the H-R diagram, BLAPs are located between hot massive main-sequence stars and hot subdwarfs. Their luminosities, $\log (L/\mbox{L}_\odot) =$ 2.2\,--\,2.6, are, however, about an order of magnitude higher than the luminosities of hot subdwarfs, while surface gravities, $\log(g/(\mbox{cm\,s}^{-2})) =$ 4.5\,--\,4.8, are much lower than the surface gravities of hot subdwarfs. No BLAP was discovered in the Magellanic Clouds \citep{2018pas6.conf..258P}, but these objects were at the limits of detection in the OGLE data.

The existence of a separate group of pulsating stars in a poorly populated region of the H-R diagram raises a question about their origin. \cite{2017NatAs...1E.166P} showed that two models reproduce pulsational characteristics of BLAPs. In the first model, a BLAP has a helium-burning core with a mass of about 1\,M$_\odot$ and can be a descendant of a main-sequence $\sim$5~M$_\odot$ star. This model was supported by the calculations of \cite{2018MNRAS.478.3871W}. In the other model, a BLAP has a degenerated helium core with a mass of $\sim$0.3~M$_\odot$ burning hydrogen in a shell. Such a configuration is typical for an evolved main-sequence star with a mass of about 1~M$_\odot$ prior to helium flash. The low-mass model was supported by \cite{2018MNRAS.477L..30R} and \cite{2018arXiv180907451C}, who proposed that BLAPs are hot counterparts of pre-extremely low-mass white dwarfs (WDs) with masses of 0.27\,--\,0.37\,M$_\odot$. This model was also favoured by \cite{2020MNRAS.492..232B}. They showed that the members of the short-period class of pre-WD pulsators discovered by \cite{2019ApJ...878L..35K} can have a similar structure as BLAPs and therefore can be deemed `high-gravity BLAPs'. Both models proposed for BLAPs require a significant part of the initial stellar mass to be stripped off during pre-BLAP evolution. This is possible in binaries \citep[e.g.,][]{2021MNRAS.507..621B}, but no BLAP was known to reside in a binary. In this context, the solution proposed by \cite{2020ApJ...903..100M} seemed to solve the problem. These authors showed that BLAPs can be explained as the surviving companions of type Ia supernovae, which begin their evolution as $\sim$3\,M$_\odot$ stars, transfer a large part of their masses to WD companions, and end up as stars with the masses of (0.76\,$\pm$\,0.1)\,M$_\odot$ at the time of supernova explosion. Either way, the origins of BLAPs remain unexplained.

We show in the present paper, based primarily on a detailed analysis of the Transiting Exoplanet Survey Satellite (TESS) photometry of HD\,133729, that this star is a binary consisting of a late-B spectral type main-sequence primary and a BLAP secondary. This makes the secondary the first known BLAP in a binary and a clue target for further studies of the properties of BLAPs and evolutionary channels leading to their origin. It is also the nearest and the brightest known star of this type.

\begin{figure*}[!ht]
\centering
\includegraphics[width=0.85\textwidth]{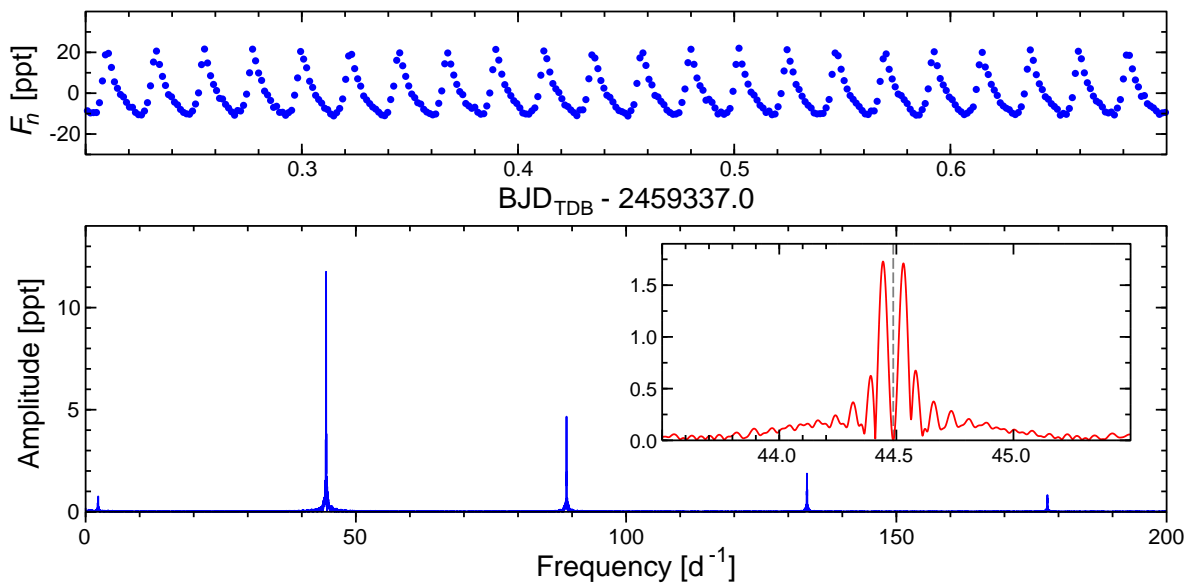}
\caption{Light curve and frequency spectrum of the TESS data of HD\,133729. Top: Half-day 2-minute cadence TESS Sector 38 light curve of HD\,133729. $F_n$ stands for the relative flux $F_i/\langle F\rangle-1$, where $F_i$ is the TESS flux for the $i$th data point and $\langle F\rangle$ is average flux from all TESS observations in Sector 38. Bottom: Frequency spectrum of the TESS Sector 38 data. In order to illustrate the occurrence of the $\pm$0.043\,d$^{-1}$ orbital sidelobes, the inset shows a 2\,d$^{-1}$-wide frequency spectrum of the residuals in the vicinity of $f_{\rm P}=44.487$\,d$^{-1}$, after subtracting $f_{\rm P}$ and its harmonics.}
\label{fig:tess}
\end{figure*}

\section{HD\,133729 and its known variability}\label{HD133729}
HD\,133729 is a bright ($V=$~9.1\,mag) star of B6/7\,V spectral type \citep{1982mcts.book.....H}. It was included in the blind search for variability among B-, A-, and F-type stars carried out by \cite{2014MNRAS.439.2078H}. They used data obtained within the Wide Angle Search for Planets (WASP) project \citep{2006PASP..118.1407P}. The search was focused on a discovery of short-period variability, mainly in rapidly-oscillating Ap (roAp) stars, pulsating Am stars, $\delta$~Scuti stars with high frequencies, and pre-main-sequence $\delta$~Scuti stars. The result of this search was the discovery of 375 stars showing variability with periods shorter than 30~minutes. The follow-up spectroscopy of 37 (about 10\%) of these high-frequency WASP pulsators, combined with the detailed analysis of their photometry, allowed \cite{2014MNRAS.439.2078H} to announce the discovery of 10 new roAp stars, 13 pulsating Am stars and $\delta$~Sct stars with frequencies higher than 100\,d$^{-1}$. 

For HD\,133729, \cite{2014MNRAS.439.2078H} reported three nearly equidistant frequencies, 88.93, 88.97, and 89.02\,d$^{-1}$, with amplitudes equal to 6, 2, and 2\,mmag, respectively. Variability with such high frequencies is not known in B-type main-sequence stars. The star was not included by these authors in the spectroscopic follow-up program and therefore its WASP photometry was not studied in detail. Hence, the nature of its variability remained unknown. The star was proposed for 2-min and 20-s cadence observations with TESS satellite, however.

The present study shows that the dominating term found by \cite{2014MNRAS.439.2078H} has a frequency equal to the lowest harmonic of the real pulsation frequency. The pulsation frequency of 44.49\,d$^{-1}$ was not detected by these authors because the lower limit of their frequency search was set to 50\,d$^{-1}$. The two other frequencies are orbital sidelobes caused by the light-travel-time effect in the system with the orbital period of about 23~days (Sect.\,\ref{sect:ltte}). 

\cite{2018AJ....155...39O} gave an upper limit for non-variability of HD\,133729 in the Kilodegree Extremely Little Telescope (KELT) data, but this study was based on integrated 30-min data, comparable with the variability period of the BLAP (Sect.\,\ref{sect:tess}). Given its low amplitude, the variability was averaged in this statistics. Some recent comments on the variability of HD\,133729 were already based on the TESS data. \cite{2020A&A...638A..59B} analysed a sample of nearly 2400 $\delta$~Sct candidates in their study aimed at finding scaling relations for $\delta$~Sct stars. From the point of view of their goals, HD\,133729 did not seem to be interesting, so they included it in the subsample of stars showing `other kind of variation'. 

\section{Analysis of photometry}
\subsection{TESS photometry}\label{sect:tess}
HD\,133729 (TIC\,75934024) was observed by the TESS satellite in Sector 11 with a 2-min cadence and in Sector 38 during TESS Extended Mission with both 2-min and 20-s cadences. The Sector 11 data were secured between 2019 April 26 and May 20, covering nearly 24~days with a 4.5-day gap in the middle of the run. The Sector 38 data were obtained between 2021 April 29 and May 26, with a short one-day gap. All TESS data were downloaded from the Barbara A.~Mikulski Archive for Space Telescopes (MAST) Portal\footnote{https://mast.stsci.edu/portal/Mashup/Clients/Mast/Portal.html}. The data consist of 13\,887 and 18\,248 data points with 2-min cadence in Sector 11 and 38, respectively. In addition, there are 109\,581 data points with 20-s cadence in Sector 38. In our analysis, we used PDCSAP fluxes, removing only some short parts at the beginning and at the end of the Sector 11 time-series. 

The TESS data from both sectors were analysed separately. Frequency spectra of the data from both sectors are fully consistent. Figure \ref{fig:tess} shows the frequency spectrum of Sector 38 2-min cadence data. The spectrum is dominated by the frequency at $f_{\rm P}=44.486$~d$^{-1}$ and its harmonics. In addition, small-amplitude signal at low ($\sim$2.2\,d$^{-1}$) frequencies can be seen (Sect.\,\ref{sect:lowfreq}). The presence of the harmonics of $f_{\rm P}$ means that the light curve is non-sinusoidal, which can be seen in the top panel of Fig.\,\ref{fig:tess}. A subtraction of $f_{\rm P}$ and its harmonics did not entirely remove the signal in the vicinity of these frequencies and some residual signal was left (Fig.\,\ref{fig:tess}, the inset in the bottom panel). The pattern was dominated by $\pm$0.043\,d$^{-1}$ sidelobes. As we show in Sect.\,\ref{sect:ltte}, the sidelobes originate from phase changes caused by the orbital motion in a binary system.

The shape of the light curve of HD\,133729 and its period equal to 32.37\,min fits perfectly the characteristics of BLAPs. The only problem is a small peak-to-peak amplitude, in TESS passband equal to about 32 parts-per-thousand (ppt). This can be explained, however, by a dilution of BLAP's light by the brighter companion, the late-B main-sequence star (Sect.\,\ref{sect:sed}). Therefore, we refer to $f_{\rm P}$ as the pulsation frequency of the BLAP. 

\subsection{Archival photometry and the changes in the pulsation period}\label{sec:ppchange}
Since BLAPs show significant period changes \citep{2017NatAs...1E.166P,2018MNRAS.478.3871W,2018arXiv180907451C}, we checked if this is also the case for HD\,133729. This can be done because archival photometry from Hipparcos space mission and two ground-based surveys, All-Sky Automated Survey \cite[ASAS,][]{1997AcA....47..467P,2000AcA....50..177P,2009ASPC..403...52P}, and WASP \citep{2006PASP..118.1407P}, is available for this star.

\begin{figure}[!ht]
\centering
\includegraphics[width=\columnwidth]{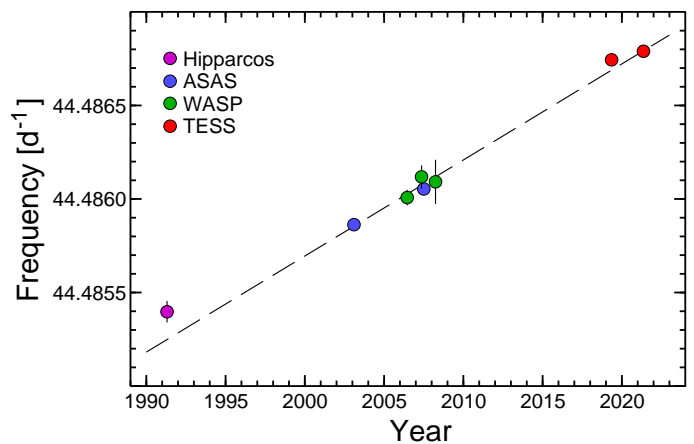}
\caption{Pulsation frequencies of the BLAP in HD\,133729 system derived from the Hipparcos, ASAS, WASP, and TESS data. Dashed line corresponds to the derived constant rate of frequency change, d$f_{\rm P}$/d$t= (1.41 \pm 0.07)\times 10^{-7}$\,d$^{-2}$.}
\label{fig:fpuls}
\end{figure}

\begin{figure}[!ht]
\centering
\includegraphics[width=\columnwidth]{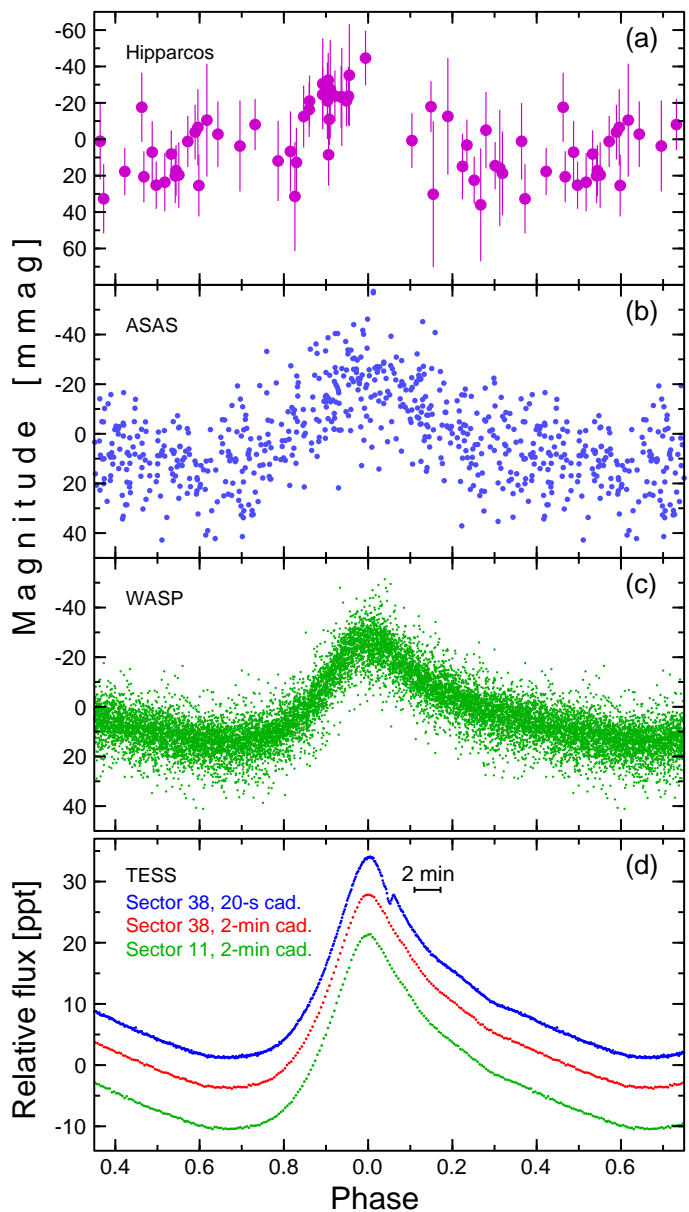}
\caption{Light curves of HD\,133729 phased with the pulsation period. (a)\,--\,(c): Hipparcos (a), ASAS (b), and WASP (c) light curves. (d) Phase-averaged TESS light curves in Sector 11 (green dots), Sector 38, 2-min cadence (red dots), and Sector 38, 20-s cadence (blue dots). For clarity, 5\,ppt offsets were applied between the light curves. The 2-min cadence data are averaged in 0.005, the 20-s cadence data, in 0.002 phase intervals. Phase 0.0 corresponds to the time of maximum light.}
\label{fig:phased-lc}
\end{figure}

The archival photometry of HD\,133729 from the Hipparcos mission, the ASAS survey, and the WASP survey, was first cleaned from outliers and some data of lower quality. Since we were interested in the light curve of the BLAP, we also removed some obvious trends in photometry by filtering out low frequencies. In order to account for phase changes due to the orbital motion, the archival time-series photometry $m(t)$ was fitted with the following variability model:
\begin{align}\label{eq:vmodel}
m(t) = m_0 &+ \sum\limits_{n=1}^{N}\{A_n\sin(2\pi nf_{\rm P}t+\phi_n)\nonumber\\ 
 &+ A_{n-}\sin[2\pi(nf_{\rm P}-f_{\rm orb})t + \phi_{n-}]\nonumber\\ 
 &+ A_{n+}\sin[2\pi(nf_{\rm P}+f_{\rm orb})t + \phi_{n+}]\}.
\end{align}
Parameters derived in this model are amplitudes $A_n$, $A_{n-}$, and $A_{n+}$, phases $\phi_n$, $\phi_{n-}$, and $\phi_{n+}$, frequencies $f_{\rm P}$ and $f_{\rm orb}$, and mean magnitude $m_0$. The value of $N$ depended on the number of harmonics of $f_{\rm P}$ detected in a given data set. We adopted $N=2$ for the Hipparcos and the ASAS data, and $N=4$ for the WASP data. For the TESS data, we had to proceed in a different way, because the data are short and each sector covers only about one orbital period. Therefore, we first derived the parameters of the orbit (Sect.\,\ref{sect:ltte}), then corrected times of observations according to the formula given by \cite{1952ApJ...116..211I} accounting for the light-travel-time delay, and only then fitted a simplified model consisting of only $f_{\rm P}$ and its harmonics up to $N=12$, that is, without the last two terms in Eq.\,(\ref{eq:vmodel}). The values of $f_{\rm P}$ derived from the fits are listed in Table \ref{tab:freq-puls} and plotted in Fig.\,\ref{fig:fpuls}. Light curves from all surveys phased with the pulsation period, are presented in Fig.\,\ref{fig:phased-lc}.
\begin{table}[!ht]
\caption{Pulsation frequencies $f_{\rm P}$ derived from fitting Eq.\,(\ref{eq:vmodel}) to the available photometric data. $N_{\rm obs}$ stands for the number of observations.}
\label{tab:freq-puls}
\centering 
\begin{tabular}{ccrrl} 
\hline\hline\noalign{\smallskip}
Source & Mean time & $N_{\rm obs}$ & $N$ & \multicolumn{1}{c}{$f_{\rm P}$} \\ 
of data & [year] & & & \multicolumn{1}{c}{[d$^{-1}$]} \\
\noalign{\smallskip}\hline\noalign{\smallskip}
Hipparcos & 1991.307 & 53 & 2 & 44.48540(6)\\
ASAS, part 1 & 2003.100 & 256 & 2 & 44.485864(17)\\
ASAS, part 2 & 2007.505 & 257 & 2 & 44.486053(17)\\
WASP, part 1 & 2006.456 & 2998 & 4 & 44.48601(4)\\
WASP, part 2 & 2007.362 & 5131 & 4 & 44.48612(6)\\
WASP, part 3 & 2008.243 & 1378 & 4 & 44.48609(12)\\
TESS, sec.\,11 & 2019.350 & 13808 & 12 & 44.486742(11)\\
TESS, sec.\,38 & 2021.363 & 17926 & 12 & 44.486792(9)\\
\noalign{\smallskip}\hline 
\end{tabular}
\end{table}

The Hipparcos photometry of HD\,133729 (HIP\,73969) is very scarce (Fig.\,\ref{fig:phased-lc}), so that detection of $f_{\rm P}$ is marginal, but it is very important from the point of view of monitoring the changes in the pulsation period, because the data extend the time span of photometric data to $\sim$30 years. The star was deemed constant in the Hipparcos catalog \citep{1997ESASP1200.....E}.

The ASAS $V$-filter data of the star (ASAS\,150658-3138.7) were downloaded from the ASAS web page\footnote{http://www.astrouw.edu.pl/asas/?page=aasc}. They cover almost nine years between December 2000 and September 2009. Therefore, we split the ASAS data into two equal parts, which resulted in two points in Fig.\,\ref{fig:fpuls}.

The WASP data of HD\,133729 were downloaded from the public database\footnote{https://wasp.cerit-sc.cz/form}. Over 11\,000 original data points were available for the star, designated as 1SWASP J150658.93-313838.9. The data span three seasons, between 2006 and 2008. For the purpose of the determination of $f_{\rm P}$, each season was analysed separately, which resulted in three points in Fig.\,\ref{fig:fpuls}.

Figure \ref{fig:fpuls} shows the resulting changes in the pulsation frequency, which at first approximation can be assumed linear. The derived rate of frequency change, d$f_{\rm P}$/d$t = (1.41 \pm 0.07)\times 10^{-7}$\,d$^{-2}$, translates into the rate of period change d$P_{\rm P}$/d$t\equiv\dot{P_{\rm P}}=(-7.11 \pm 0.33)\times 10^{-11}$ or $r \equiv \dot{P_{\rm P}}/P_{\rm P} = (-11.5 \pm 0.6)\times 10^{-7}$\,(yr)$^{-1}$, where $P_{\rm P}=f_{\rm P}^{-1}$ is the pulsation period. This is the largest negative value of the rate of the period change of all known BLAPs, if the uncertain value of $\dot{P_{\rm P}}$ for OGLE-BLAP-002 derived by \cite{2018MNRAS.478.3871W}, is not taken into account. 

\subsection{Light-travel-time effect}\label{sect:ltte}
A periodic photometric signal sent by a component of a binary can serve as a clock and be used to derive the same orbital parameters that are derived for a single-lined spectroscopic binary. The involved effect is usually called light-travel-time effect (hereafter LTTE) and results in the occurrence of orbital sidelobes in frequency spectra at a separation equal to an orbital frequency, $f_{\rm orb}$. For HD\,133729, a separation of orbital sidelobes equal to about 0.043\,d$^{-1}$ (Fig.\,\ref{fig:tess}) indicates for an orbital period of $P_{\rm orb}=f_{\rm orb}^{-1}\approx 23$~days.

In order to analyze LTTE, we used times of maximum light of the pulsation of the BLAP (Fig.\,\ref{fig:phased-lc}). Since data used to derive a single time of maximum light have to cover a reasonably small part of the orbital period, the only data suitable for this kind of analysis are the TESS and the WASP data. The TESS data were split into $\sim$1-day long subsets. This resulted in 45 subsets, 20 for Sector 11 and 25 for Sector 38 observations. Then, a smoothed light curve was obtained by phasing Sector 38 2-min cadence TESS light curve (Fig.\,\ref{fig:phased-lc}). Next, the smoothed light curve was fitted to each subset separately by means of the least squares. In effect, a single time of maximum light was obtained for each subset. It was the closest to the mid-time of observations in a given subset. A similar procedure was applied to the WASP data, but in this case, we chose subsets corresponding to individual nights. Nights with less than 20 data points were excluded. In this way, 117 times of maximum light for the WASP data were obtained. The times of observations in the WASP data are given in Heliocentric Julian Days (HJD), while for the TESS data, in Barycentric Julian Days in the standard of Barycentric Dynamical Time, TDB (BJD$_{\rm TDB}$). In order to make them consistent, we converted the WASP times of maximum light to BJD$_{\rm TDB}$ using the on-line tool\footnote{https://astroutils.astronomy.osu.edu/time/hjd2bjd.html} provided by J.\,Eastman \citep{2010PASP..122..935E}. All times of maximum light, $T_{\rm max}$, are given in Table \ref{tab:tmax} in Appendix \ref{app:tmax}.

Figure \ref{fig:phased-lc} shows the superb quality of the TESS data. It can be also seen that the average light curves for 2-min cadence Sector 11 and Sector 38 data match perfectly. Since 2-min cadence TESS data are sufficient for the $O-C$ analysis, we did not use 20-s cadence data for this purpose. However, these data were analysed in the same way as 2-min cadence data and, for completeness, the phase-averaged light curve for these data is shown in Fig.\,\ref{fig:phased-lc}d. In general, the 20-s cadence data are consistent with the 2-min cadence data. The only difference is an unusual tiny drop of flux lasting only about 40\,s at phase 0.05, which can be seen in the light curve of the 20-s cadence data. The feature is averaged in 2-min cadence data. It may have the same origin as the small drops following the epoch of maximum light that can be seen in the light curves of OGLE-BLAP-009 \citep{2017NatAs...1E.166P,2020MNRAS.496.1105M} and ZGP-BLAP-05, -06, and -07 \citep{2022MNRAS.511.4971M}.  

\begin{figure}[!ht]
\centering
\includegraphics[width=\columnwidth]{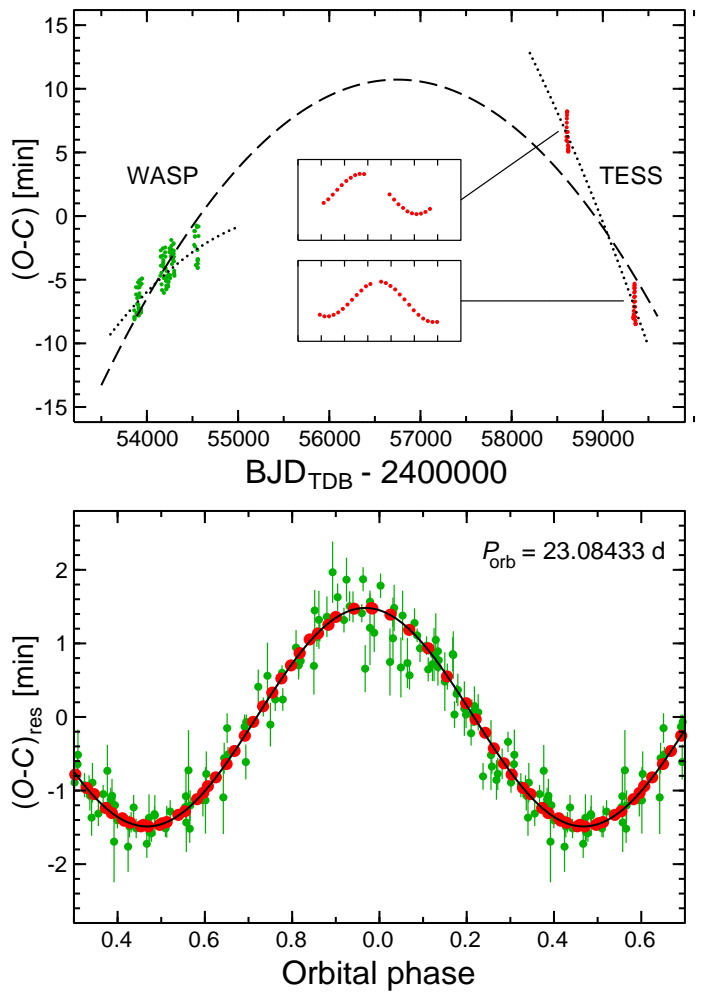}
\caption{$O-C$ diagrams for the pulsation period of HD\,133729. Top: $O-C$ diagram for WASP (green dots) and TESS (red dots) times of maximum light obtained using Eq.\,(\ref{eq:ephem}) as a reference. Dashed line shows fit with the assumed period change d$P$/d$t =-7.11\times 10^{-11}$ derived in Sect.\,\ref{sec:ppchange}. Dotted lines show fits of parabolae with the same d$P$/d$t$ separately to WASP and TESS data. Ordinate ranges one pulsation period. The two insets show zoomed TESS data, the abscissa ticks are separated by 5\,d. Bottom: Residuals from the fits shown in the upper panel phased with the orbital period of 23.08433\,d. Red and green dots correspond to TESS and WASP data, respectively. Solid black line is a fit of Eq.\,(\ref{eq:oc}) to the residual $O-C$ data. Parameters of the fit are given in Table \ref{tab:params}. Phase 0.0 corresponds to $T_0$.}
\label{fig:oc}
\end{figure}

Having obtained times of maximum light, $T_{\rm max}$, we used them to obtain $O-C$ diagram shown in Fig.\,\ref{fig:oc}, upper panel. The values of $O-C$ were calculated with respect to the following ephemeris:
\begin{align}\label{eq:ephem}
C(E) = &T_{\rm max}^C(E) =\nonumber\\&\mbox{BJD}_{\rm TDB}\,2\,453\,860.487 + 0.022478855 \times E,
\end{align}
where $E$ is the number of pulsation periods elapsed from the initial epoch. The values of $O-C = T_{\rm max} - T_{\rm max}^C$, where $T_{\rm max}$ and $T_{\rm max}^C$ stand for the observed and calculated times of maximum light, respectively. They are given in Table \ref{tab:tmax}. The $O-C$ diagram shows periodic variation with the period of about 23~d superimposed on changes on a much longer time scale. As expected, the changes on a long-time scale reflect, to some extent, the decrease of pulsation period suggested in Sect.\,\ref{sec:ppchange}, but it is also obvious that a constant rate of period change, represented by the dashed-line parabola, does not fit the observations well. The real changes in the pulsation period are more complicated. Unfortunately, they cannot be traced by the available photometry because of the gaps in the data. Since we are interested in the periodic variation in the $O-C$ diagram, we subtracted the long term seen in Fig.\,\ref{fig:oc}, upper panel, separately for the WASP and the TESS data, by fitting `locally' and subtracting parabolae with a quadratic term corresponding to $\dot{P}=-7.11\times 10^{-11}$ (Sect.\,\ref{sec:ppchange}). These parabolae are shown with the dotted lines in Fig.\,\ref{fig:oc}.

Residual $O-C$ values, $(O-C)_{\rm res}$, were used to derive parameters of the orbit of the BLAP by fitting equation
\begin{equation}
\label{eq:oc}
(O-C)_{\rm res}(t) = \tau(a_2\sin i, P_{\rm orb}, T_0, e, \omega, t) + \mbox{const}
\end{equation}
to the residual data. The values of $(O-C)_{\rm res}$ are also given in Table \ref{tab:tmax}. In Eq.\,(\ref{eq:oc}), the light-travel time $\tau$ is defined by eq.\,(3) of \cite{1952ApJ...116..211I}. Derived parameters are listed in Table \ref{tab:params}. The residual $O-C$ values phased with the derived orbital period are shown in the bottom panel of Fig.\,\ref{fig:oc}. The orbit is marginally eccentric. Although the WASP data resulted in much less precise times of maximum light than the TESS data, they are important for constraining $P_{\rm orb}$, because they increase the span of data to about 240 orbital cycles. High precision of $K_2$ and mass function $f(M_1)$\footnote{We consequently use subscript `1' and name `primary' for the late B-type component, and subscript `2' and name `secondary' for the BLAP.} is the result of the precise determination of the amplitude of LTTE, a consequence of the high precision of the times of maximum light derived from the TESS data. The standard deviation of residuals from the fit shown in Fig.\,\ref{fig:oc} amounts to 16.0~s for the WASP data and only 0.7~s for the TESS data.
\begin{table}[!ht]
\caption{Parameters of the orbit of BLAP derived from fitting the $O-C$ diagram.}
\label{tab:params}
\centering 
\begin{tabular}{lc} 
\hline\hline\noalign{\smallskip}
Parameter & Value \\ 
\noalign{\smallskip}\hline\noalign{\smallskip}
Orbital period, $P_{\rm orb}$ & 23.08433 $\pm$ 0.00023 d\\
Time of periastron passage, $T_0$ & BJD$_{\rm TDB}$\,2\,453\,900.4 $\pm$ 1.2\\
Eccentricity, $e$ & 0.0055 $\pm$ 0.0017\\
Argument of periastron, $\omega$ & 102 $\pm$ 19$\degr$ \\
Projected major semiaxis of the& \\
~~BLAP's absolute orbit, $a_2\sin i$ & 38.389 $\pm$ 0.032 R$_\odot$\\
Half-range of the BLAP's &\\
~~radial-velocity variation, $K_2$ & 84.13 $\pm$ 0.07 km\,s$^{-1}$\\
Mass function, $f(M_1)$ & 1.424 $\pm$ 0.004 M$_\odot$\\
\noalign{\smallskip}\hline 
\end{tabular}
\end{table}

\subsection{Low frequencies}\label{sect:lowfreq}
\begin{figure}[!ht]
\centering
\includegraphics[width=\columnwidth]{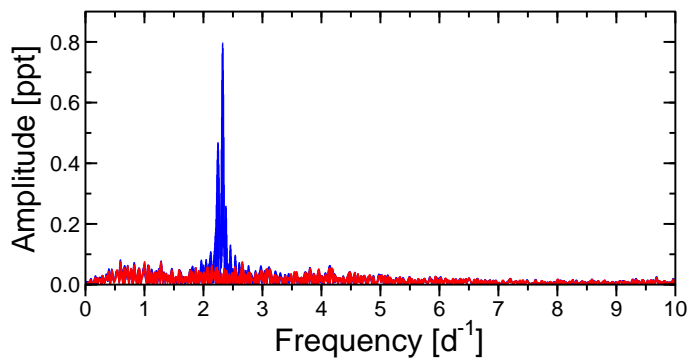}
\caption{Frequency spectrum of the combined sectors of the TESS data after subtracting BLAP pulsations (blue), and after prewhitening four periodic terms that were detected (red). More details are given in the text.}
\label{fig:lf}
\end{figure}

In addition to the variability caused by the pulsations of the BLAP, a weak signal at frequencies close to 2.2\,d$^{-1}$ was detected in the TESS data. The signal can be seen even in Fig.\,\ref{fig:tess}. We have combined data from both TESS sectors subtracting first the BLAP pulsations. The frequency spectrum of these data is shown in Fig.\,\ref{fig:lf}. The strongest signal can be well represented by four periodic terms with frequencies 2.3266, 2.2431, 2.2104, and 2.2876\,d$^{-1}$ and semi-amplitudes of 0.76, 0.41, 0.15, and 0.11\,ppt, respectively. Due to the large gap between both sectors, the frequencies are subject to strong aliasing. We do not know which component is responsible for this variability. Given the primary is a late B-type star, a suitable explanation is that this component is a slowly pulsating B-type (SPB) star, but in general, we cannot exclude the possibility that the low-frequency variability originates in the BLAP.

\section{SED fitting and the parameters of the components}\label{sect:sed}
\begin{figure}[!ht]
\centering
\includegraphics[width=\columnwidth]{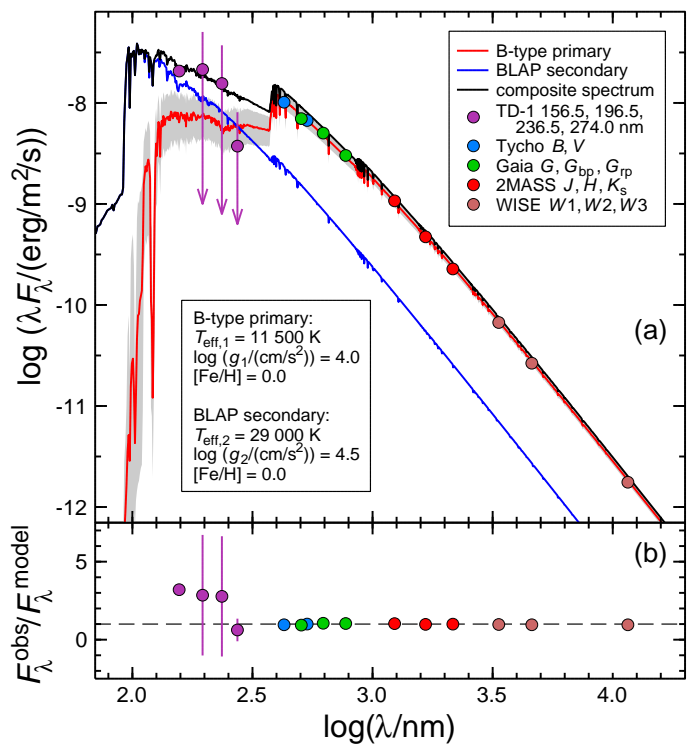}
\caption{Results of fitting SED to the photometric data of HD\,133729. (a) Colour-coded points with error bars stand for observed fluxes de-reddened by the VOSA tool. The spectrum fitted to the optical and IR data is shown with solid red curve. The shaded area marks the effect of $\pm 1000$\,K uncertainty in $T_{\rm eff,1}$. The adopted spectrum of the BLAP scaled to account for the UV data is marked with blue line. (b) Residuals from fitting SED to the visual and IR photometry of HD\,133729 in the terms of flux ratio.}
\label{fig:sed}
\end{figure}

In order to verify if multi-band photometry of HD\,133729 is consistent with a system consisting of a late B-type star and a BLAP, we performed spectral energy distribution (SED) analysis using VOSA\footnote{http://svo2.cab.inta-csic.es/theory/vosa/} on-line tool \citep{2008A&A...492..277B}. In the optical part of the spectrum, we used Tycho $B_{\rm T}$ and $V_{\rm T}$ \citep{2000A&A...355L..27H} and Gaia $G$, $G_{\rm bp}$, and $G_{\rm rp}$ \citep{2018A&A...616A...1G} photometry. In the infrared (IR) domain, we chose photometry in 2MASS $JHK_{\rm s}$ \citep{2003yCat.2246....0C}, and WISE $W1$, $W2$, and $W3$ \citep{2012yCat.2311....0C} passbands. Assuming that contribution of BLAP in the optical and IR domains is small compared to the contribution of the B-type primary, we obtained a best-fitting model with effective temperature $T_{\rm eff,1}=11\,500$\,K and surface gravity $\log (g_1/(\mbox{cm\,s}^{-2}))=4.0$\,dex. We assumed total absorption in Johnson $V$ band $A_V=0.386$\,mag. This value was taken from Gaia-2MASS three-dimensional map of absorption presented by \cite{2019A&A...625A.135L} using their on-line tool\footnote{https://astro.acri-st.fr/gaia\_dev/}. During standard least-squares fitting procedure, we used Castelli-Kurucz ODFNEW/NOVER synthetic spectra \citep{1997A&A...318..841C} and assumed solar metallicity. The fit is shown in Fig.\,\ref{fig:sed}a, the residuals, in Fig.\,\ref{fig:sed}b. The plots show also another source of photometric data, ultraviolet (UV) fluxes collected by Thor-Delta 1A (TD-1) satellite in four passbands, 156.5, 196.5, 236.5 and 274.0\,nm \citep{1978csuf.book.....T,1995yCat.2059....0T}. They were not used in the fit because contribution of the BLAP in the UV is significant. The star is barely detected in the three longest-wavelength TD-1 passbands (the errors are larger than the measured fluxes), but in the 156.5-nm band the flux amounts to (5.25\,$\pm$\,0.49)\,$\times$\,10$^{-14}$\,W\,m$^{-2}$\,(nm)$^{-1}$. A comparison of the ratios of TD-1 fluxes to those of the fitted SED model shows (Fig.\,\ref{fig:sed}b) that the observed flux in the 156.5-nm band is roughly three times higher than expected from a late B-type star. This can be explained only by the presence of a hot secondary, the BLAP. We therefore took a spectrum with parameters representing known BLAPs, $T_{\rm eff,2}=29\,000$\,K and $\log (g_2/(\mbox{cm\,s}^{-2}))=4.5$\,dex, and scaled it to obtain a best fit of the combined spectrum to the TD-1 observations. The result is shown in Fig.\,\ref{fig:sed}a with a black line. It can be seen that by adding a hot companion to the fit, we obtained a satisfactory agreement in the whole range of the available passbands.

The fitted $T_{\rm eff,1}$ corresponds to spectral type B8 or even B9, which is slightly later than B6/7\,V derived by \cite{1982mcts.book.....H}. Since secondary's contribution to the total flux in the blue region of the spectrum exceeds 20\%, we speculate that secondary's lines could have affected this classification.

The SEDs of both components allowed us to estimate a relative contribution of the components in different passbands. In particular, the BLAP contribution to the total flux of the system amounts to about 19\% in Johnson $V$ and 16\% in the TESS band. The latter value allows to estimate the intrinsic amplitude of the BLAP. The observed peak-to-peak amplitude of the BLAP in the TESS band amounts to 31.6\,ppt, which translates into the intrinsic amplitude of 0.21\,mag, in full agreement with the $I$-filter amplitudes of BLAPs derived by \cite{2017NatAs...1E.166P}.

\begin{figure}[!ht]
\centering
\includegraphics[width=\columnwidth]{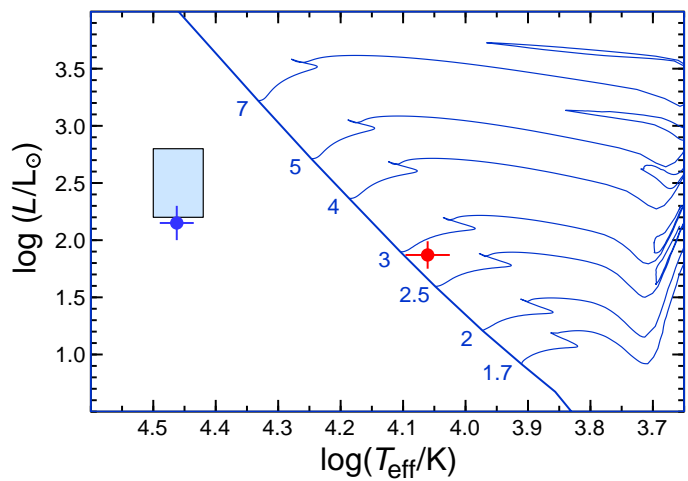}
\caption{Positions of the components of HD\,133729 in the H-R diagram. The B-type primary is marked with the red dot, the BLAP, with the blue dot. Light blue rectangle marks the the position of BLAPs as defined by \cite{2017NatAs...1E.166P}. Evolutionary tracks, labeled with mass in M$_\odot$, and zero-age main sequence line were taken from non-rotating models of \cite{2012A&A...537A.146E}.}
\label{fig:hr}
\end{figure}

In order to locate the components of the system in the H-R diagram, we used the results of the SED fitting. The measured Gaia Early Data Release 3 parallax of HD\,133729 amounts to 2.136\,$\pm$\,0.029\,mas, which translates into the distance of $D=463\,\pm\,7$\,pc \citep{2021AJ....161..147B}. We calculated $V$ magnitude of the system translating Tycho $B_{\rm T}$ and $V_{\rm T}$ magnitudes by means of eq.\,(1) of \cite{1998A&A...333..673O}. It is equal to 9.174\,mag. In consequence, we obtained $V_1\approx 9.37$\,mag and $V_2\approx 11.0$\,mag adopting the relative contributions of the components estimated from the SED fitting. This resulted in the absolute magnitude of the primary equal to $M_{V,1}=+0.66$\,mag. Adopting bolometric correction (BC)$_{V,1}=-0.596$\,mag, calculated from the equation provided by \cite{2020MNRAS.495.2738P} for Johnson $V$ band, assuming $T_{\rm eff,1}=11\,500$\,K and $M_{{\rm bol},\odot}=4.74$\,mag, we obtained the bolometric magnitude of the primary $M_{\rm bol,1}=+0.06$\,mag and its luminosity $\log (L_1/\mbox{L}_\odot)=1.87\,\pm\,0.12$. In these calculations, an uncertainty of $\pm$1000\,K for $T_{\rm eff,1}$ was adopted. As can be seen in Fig.\,\ref{fig:sed}a, this value encompasses all photometric data and small differences in the visual and IR domains between the spectrum of the primary and the combined spectrum. Uncertainty of $\log (L_1/\mbox{L}_\odot)$ depends on the uncertainties of four parameters: distance, $V_1$, $A_V$, and (BC)$_{V,1}$. Adopting $\sigma(A_V)=0.1$\,mag, $\sigma(V_1)=0.15$\,mag, and $\sigma((\mbox{BC})_{V,1})=0.25$\,mag, we obtained the uncertainty of $\log (L_1/\mbox{L}_\odot)$ equal to 0.12\,dex. A comparison of the location of the primary with the evolutionary models (Fig.\,\ref{fig:hr}) places the star in the main sequence phase of evolution, as expected from its luminosity class. The inferred mass of the primary amounts to (2.85\,$\pm$\,0.25)\,M$_\odot$. 

The SED fitting allowed us to estimate relative luminosities of the components by integrating the SEDs of both components. A relative luminosity, $L_2/L_1 \approx 1.9$, combined with the luminosity of the primary derived above gives $\log(L_2/\mbox{L}_\odot)=2.15\,\pm\,0.15$. Along with the secondary's effective temperature adopted for SED fitting ($T_{\rm eff,2} = 29\,000$\,K), this allows to plot the position of the secondary in the H-R diagram. As can be seen in Fig.\,\ref{fig:hr}, it fits within the errors the `instability strip' of BLAPs as defined by \cite{2017NatAs...1E.166P}. The location and size of the strip is still rather uncertain. It may extend towards lower luminosities, especially if BLAPs and high-gravity BLAPs form a homogeneous group. The location of the secondary in the H-R diagram is another argument in favour of the recognition of the star as a BLAP. 

\section{Discussion and conclusions}
BLAPs are very rare objects. Only 14 were discovered during the extensive survey of the OGLE data by \cite{2017NatAs...1E.166P} and \cite{2018pas6.conf..258P}. Recently, \cite{2022MNRAS.511.4971M} announced discovery of a sample of 16 (candidate) BLAPs and 6 candidate high-gravity BLAPs using Zwicky Transient Facility \citep{2019PASP..131a8002B}  DR3 time-series photometry. Adding four high-gravity BLAPs found by \cite{2019ApJ...878L..35K} and one candidate with the period of 18.9\,min found recently by \cite{2022MNRAS.509.2362L}, the secondary of HD\,133729 is a star belonging to the class which presently counts less than 30 members. This star is unusual for several reasons. First, because it is the only known BLAP that resides in a binary. The binary is wide enough to make use of the light-travel-time effect for the determination of precise spectroscopic elements of its orbit (Sect.\,\ref{sect:ltte}). Next, it is the nearest and the brightest star of this type. With the estimated $V$ magnitude of about 11.0 (Sect.\,\ref{sect:sed}), it is almost 5\,mag brighter than the next brightest BLAP, OGLE-BLAP-009 ($V\approx 15.7$\,mag). Finally, the unusual short-living feature in the descending branch of the light curve (Fig.\,\ref{fig:phased-lc}) is exceptional among radially pulsating stars. It was detected only due to a very good time resolution in the Sector 38 20-s cadence TESS data.

In an obvious way, the discovery of a BLAP in a bright binary system prompts for a detailed follow-up spectroscopic study, we are going to carry out. First of all, this should provide the mass ratio of the components. Given the relatively tight constraint on the primary's mass (Sect.\,\ref{sect:sed}), with the relative error of the order of 10\%, this should allow to constrain $M_2$ with a comparable precision. For this purpose, we need to measure precisely $K_1$, the half-range of the primary's radial-velocity variation. In this context, the domination of the primary in the optical part of the spectrum of the system might be a favourable circumstance. We already know $K_2$ with a 0.1\% precision from the $O-C$ analysis (Sect.\,\ref{sect:ltte}). Such precision of $K_2$ would be hardly achievable from the analysis of optical spectra, as the secondary contributes only 15\,--\,20\% to the optical spectrum and its spectral lines would be affected by the presence of the short-period large-amplitude pulsations. A precise value of the mass ratio, allowing to constrain also the mass of the secondary, should allow to conclude on the evolutionary status of this BLAP and choose at least between scenarios leading to a shell-burning star with a mass of 0.3\,--\,0.4\,M$_\odot$ and those that predict a core-helium-burning star with a higher mass of 0.5\,--\,1.0\,M$_\odot$.

If a coevality of both components is assumed and no mass transfer onto the primary took place, the progenitor of the BLAP in the HD\,133729 system must have started its evolution with an initial mass larger than $\sim$3\,M$_\odot$. This would exclude evolutionary scenarios that lead to low-mass BLAPs, because they start with $\sim$1\,M$_\odot$ progenitors. A detailed discussion of the possible evolutionary scenarios of the BLAP in HD\,133729 system is, however, beyond the scope of this paper, and is postponed to the spectroscopic study.

\begin{acknowledgements} 
Publication of this paper was partially financed by the University of Wroc{\l}aw program `Initiative of Excellence -- Research University'. AP, KK, and PKS were supported by the National Science Centre (NCN) grants no.~2016/21/B/ST9/01126, 2019/33/N/ST9/01818, and 2019/35/N/ST9/03805, respectively. This work used the SIMBAD and Vizier services operated by Centre des Donn\'ees astronomiques de Strasbourg (France), and bibliographic references from the Astrophysics Data System maintained by SAO/NASA. This paper makes use of data from the DR1 of the WASP data \citep{2010A&A...520L..10B} as provided by the WASP consortium, and the computing and storage facilities at the CERIT Scientific Cloud, reg.\,no.\,CZ.1.05/3.2.00/08.0144 which is operated by Masaryk University, Czech Republic. This publication makes use of VOSA, developed under the Spanish Virtual Observatory project supported from the Spanish MICINN through grant AyA2008-02156. This work has made use of data from the European Space Agency (ESA) mission {\it Gaia}(\footnote{https://www.cosmos.esa.int/gaia}, processed by the {\it Gaia} Data Processing and Analysis Consortium (DPAC\footnote{https://www.cosmos.esa.int/web/gaia/dpac/consortium}). Funding for the DPAC has been provided by national institutions, in particular the institutions participating in the {\it Gaia} Multilateral Agreement.
\end{acknowledgements}
\bibliography{BLAP}

\begin{appendix}
\section{Table with the values of $T_{\rm max}$}\label{app:tmax}
\onecolumn
\tiny
\begin{longtable}{lrrrlrrr}
\caption{\label{tab:tmax}Times of maximum light obtained using the WASP data and the TESS 2-min cadence data. All times are given as BJD$_{\rm TDB} -$ 2\,450\,000.0. The values of $E$, $(O-C)$ and $(O-C)_{\rm res}$ are explained in Sect.\,\ref{sect:ltte}.}\\
\hline\hline
\multicolumn{1}{c}{$T_{\rm max}$} & \multicolumn{1}{c}{$E$} & \multicolumn{1}{c}{$(O-C)$} & \multicolumn{1}{c}{$(O-C)_{\rm res}$} & \multicolumn{1}{c}{$T_{\rm max}$} & \multicolumn{1}{c}{$E$} & \multicolumn{1}{c}{$(O-C)$} & \multicolumn{1}{c}{$(O-C)_{\rm res}$}\\
&  & \multicolumn{1}{c}{[d]} & \multicolumn{1}{c}{[d]} & & & \multicolumn{1}{c}{[d]} & \multicolumn{1}{c}{[d]}\\
\endfirsthead
\caption{continued.}\\
\hline\hline
\multicolumn{1}{c}{$T_{\rm max}$} & \multicolumn{1}{c}{$E$} & \multicolumn{1}{c}{$(O-C)$} & \multicolumn{1}{c}{$(O-C)_{\rm res}$} & \multicolumn{1}{c}{$T_{\rm max}$} & \multicolumn{1}{c}{$E$} & \multicolumn{1}{c}{$(O-C)$} & \multicolumn{1}{c}{$(O-C)_{\rm res}$}\\
&  & \multicolumn{1}{c}{[d]} & \multicolumn{1}{c}{[d]} & & & \multicolumn{1}{c}{[d]} & \multicolumn{1}{c}{[d]}\\
\hline
\endhead
\hline
\endfoot
 3860.48186(15) & 0 & $-$0.00514 & $-$0.00054 & 4270.29687(31) & 18231 & $-$0.00214 & 0.00052 \\
 3862.43715(11) & 87 & $-$0.00551 & $-$0.00091 & 4271.33077(13) & 18277 & $-$0.00226 & 0.00039 \\
 3863.44864(7) & 132 & $-$0.00557 & $-$0.00097 & 4272.31990(11) & 18321 & $-$0.00220 & 0.00045 \\
 3864.43764(9) & 176 & $-$0.00564 & $-$0.00105 & 4273.33139(12) & 18366 & $-$0.00227 & 0.00038 \\
 3865.40437(7) & 219 & $-$0.00550 & $-$0.00092 & 4274.32010(12) & 18410 & $-$0.00262 & 0.00002 \\
 3880.39939(9) & 886 & $-$0.00388 & 0.00062 & 4286.41404(16) & 18948 & $-$0.00231 & 0.00028 \\
 3881.38806(8) & 930 & $-$0.00428 & 0.00022 & 4287.31307(10) & 18988 & $-$0.00242 & 0.00016 \\
 3886.39982(9) & 1153 & $-$0.00530 & $-$0.00083 & 4291.24777(14) & 19163 & $-$0.00153 & 0.00104 \\
 3887.38887(11) & 1197 & $-$0.00532 & $-$0.00085 & 4292.32680(7) & 19211 & $-$0.00148 & 0.00109 \\
 3890.37860(14) & 1330 & $-$0.00528 & $-$0.00084 & 4294.30437(23) & 19299 & $-$0.00205 & 0.00051 \\
 3892.37971(6) & 1419 & $-$0.00479 & $-$0.00035 & 4295.31609(8) & 19344 & $-$0.00188 & 0.00067 \\
 3893.34661(7) & 1462 & $-$0.00448 & $-$0.00005 & 4296.28234(27) & 19387 & $-$0.00222 & 0.00033 \\
 3902.38397(6) & 1864 & $-$0.00361 & 0.00077 & 4297.31612(25) & 19433 & $-$0.00247 & 0.00008 \\
 3903.48514(17) & 1913 & $-$0.00391 & 0.00047 & 4298.30456(12) & 19477 & $-$0.00310 & $-$0.00056 \\
 3904.36138(12) & 1952 & $-$0.00435 & 0.00002 & 4299.31623(14) & 19522 & $-$0.00298 & $-$0.00044 \\
 3905.35044(9) & 1996 & $-$0.00436 & 0.00001 & 4300.30510(11) & 19566 & $-$0.00318 & $-$0.00064 \\
 3906.33903(6) & 2040 & $-$0.00483 & $-$0.00047 & 4301.29409(8) & 19610 & $-$0.00326 & $-$0.00073 \\
 3907.35044(9) & 2085 & $-$0.00498 & $-$0.00062 & 4518.55391(10) & 29275 & $-$0.00157 & 0.00016 \\
 3909.32846(15) & 2173 & $-$0.00509 & $-$0.00074 & 4520.46537(27) & 29360 & $-$0.00081 & 0.00092 \\
 3910.33976(9) & 2218 & $-$0.00534 & $-$0.00099 & 4525.52310(13) & 29585 & $-$0.00082 & 0.00089 \\
 3911.32874(8) & 2262 & $-$0.00543 & $-$0.00109 & 4526.55684(25) & 29631 & $-$0.00112 & 0.00059 \\
 3913.23980(25) & 2347 & $-$0.00507 & $-$0.00075 & 4527.54591(16) & 29675 & $-$0.00111 & 0.00059 \\
 3916.34255(11) & 2485 & $-$0.00440 & $-$0.00009 & 4528.55691(12) & 29720 & $-$0.00166 & 0.00004 \\
 3917.33181(9) & 2529 & $-$0.00421 & 0.00009 & 4536.53590(27) & 30075 & $-$0.00267 & $-$0.00100 \\
 3918.32121(7) & 2573 & $-$0.00388 & 0.00042 & 4544.54052(12) & 30431 & $-$0.00052 & 0.00113 \\
 3919.31040(6) & 2617 & $-$0.00377 & 0.00053 & 4547.53012(21) & 30564 & $-$0.00061 & 0.00103 \\
 3920.32222(8) & 2662 & $-$0.00349 & 0.00080 & 4553.55320(9) & 30832 & $-$0.00185 & $-$0.00023 \\
 3926.27885(15) & 2927 & $-$0.00376 & 0.00050 & 4554.51929(14) & 30875 & $-$0.00236 & $-$0.00074 \\
 3927.38040(22) & 2976 & $-$0.00367 & 0.00059 & 4555.53076(7) & 30920 & $-$0.00244 & $-$0.00083 \\
 3928.32378(11) & 3018 & $-$0.00440 & $-$0.00015 & 4556.54191(23) & 30965 & $-$0.00283 & $-$0.00122 \\
 3934.36983(22) & 3287 & $-$0.00517 & $-$0.00095 & 4557.53101(13) & 31009 & $-$0.00280 & $-$0.00120 \\
 3935.24641(8) & 3326 & $-$0.00526 & $-$0.00105 & 4558.58773(14) & 31056 & $-$0.00259 & $-$0.00099 \\
 3936.28064(6) & 3372 & $-$0.00506 & $-$0.00085 & 4560.49864(15) & 31141 & $-$0.00238 & $-$0.00079 \\
 3942.28386(10) & 3639 & $-$0.00369 & 0.00049 & 4561.57766(34) & 31189 & $-$0.00235 & $-$0.00076 \\
 3943.25073(9) & 3682 & $-$0.00342 & 0.00075 & 4565.55768(13) & 31366 & $-$0.00109 & 0.00049 \\
 3947.27445(21) & 3861 & $-$0.00341 & 0.00074 & 4566.41239(19) & 31404 & $-$0.00057 & 0.00101 \\
 4148.55027(20) & 12815 & $-$0.00326 & $-$0.00007 & 8600.517360(9) & 210866 & 0.004122 & $-$0.000653 \\
 4151.54099(19) & 12948 & $-$0.00222 & 0.00095 & 8601.506628(9) & 210910 & 0.004320 & $-$0.000444 \\
 4152.57537(20) & 12994 & $-$0.00187 & 0.00130 & 8602.473475(8) & 210953 & 0.004576 & $-$0.000177 \\
 4153.56361(22) & 13038 & $-$0.00270 & 0.00046 & 8603.440333(9) & 210996 & 0.004843 & 0.000101 \\
 4155.54226(23) & 13126 & $-$0.00219 & 0.00096 & 8604.407175(9) & 211039 & 0.005095 & 0.000364 \\
 4159.54260(17) & 13304 & $-$0.00309 & 0.00004 & 8605.373993(8) & 211082 & 0.005322 & 0.000603 \\
 4160.50856(18) & 13347 & $-$0.00372 & $-$0.00059 & 8606.340758(8) & 211125 & 0.005496 & 0.000788 \\
 4165.52103(19) & 13570 & $-$0.00404 & $-$0.00093 & 8607.285015(8) & 211167 & 0.005641 & 0.000944 \\
 4171.47927(31) & 13835 & $-$0.00269 & 0.00039 & 8608.251673(7) & 211210 & 0.005708 & 0.001022 \\
 4175.52600(11) & 14015 & $-$0.00215 & 0.00092 & 8609.218254(7) & 211253 & 0.005699 & 0.001024 \\
 4176.53794(11) & 14060 & $-$0.00176 & 0.00130 & 8614.679507(9) & 211496 & 0.004590 & $-$0.000022 \\
 4178.53774(28) & 14149 & $-$0.00258 & 0.00047 & 8615.645813(8) & 211539 & 0.004305 & $-$0.000296 \\
 4179.54947(12) & 14194 & $-$0.00240 & 0.00065 & 8616.589667(7) & 211581 & 0.004047 & $-$0.000543 \\
 4180.53843(10) & 14238 & $-$0.00251 & 0.00053 & 8617.578539(8) & 211625 & 0.003850 & $-$0.000728 \\
 4181.54970(15) & 14283 & $-$0.00278 & 0.00026 & 8618.522462(7) & 211667 & 0.003660 & $-$0.000907 \\
 4184.53879(23) & 14416 & $-$0.00338 & $-$0.00036 & 8619.488950(7) & 211710 & 0.003558 & $-$0.000998 \\
 4186.44868(38) & 14501 & $-$0.00419 & $-$0.00118 & 8620.455495(8) & 211753 & 0.003512 & $-$0.001033 \\
 4199.55607(8) & 15084 & $-$0.00198 & 0.00098 & 8621.422116(7) & 211796 & 0.003542 & $-$0.000992 \\
 4200.54540(11) & 15128 & $-$0.00171 & 0.00124 & 8622.388797(7) & 211839 & 0.003633 & $-$0.000890 \\
 4203.46716(25) & 15258 & $-$0.00221 & 0.00073 & 8623.355550(7) & 211882 & 0.003795 & $-$0.000716 \\
 4205.51212(18) & 15349 & $-$0.00283 & 0.00010 & 9334.689629(7) & 243527 & $-$0.005492 & $-$0.000956 \\
 4206.38833(33) & 15388 & $-$0.00329 & $-$0.00036 & 9335.678610(7) & 243571 & $-$0.005582 & $-$0.001032 \\
 4207.60211(10) & 15442 & $-$0.00337 & $-$0.00045 & 9336.667680(7) & 243615 & $-$0.005581 & $-$0.001018 \\
 4208.38870(18) & 15477 & $-$0.00354 & $-$0.00062 & 9337.656829(7) & 243659 & $-$0.005501 & $-$0.000924 \\
 4211.51280(8) & 15616 & $-$0.00400 & $-$0.00110 & 9338.646033(7) & 243703 & $-$0.005367 & $-$0.000776 \\
 4212.50207(10) & 15660 & $-$0.00379 & $-$0.00089 & 9339.635295(7) & 243747 & $-$0.005175 & $-$0.000570 \\
 4213.53595(13) & 15706 & $-$0.00395 & $-$0.00105 & 9340.624600(7) & 243791 & $-$0.004939 & $-$0.000321 \\
 4214.41314(21) & 15745 & $-$0.00343 & $-$0.00054 & 9341.613931(7) & 243835 & $-$0.004678 & $-$0.000046 \\
 4215.51504(14) & 15794 & $-$0.00299 & $-$0.00011 & 9342.625741(8) & 243880 & $-$0.004416 & 0.000230 \\
 4216.50379(16) & 15838 & $-$0.00331 & $-$0.00042 & 9343.615054(8) & 243924 & $-$0.004173 & 0.000487 \\
 4231.42930(17) & 16502 & $-$0.00377 & $-$0.00095 & 9344.604356(8) & 243968 & $-$0.003941 & 0.000732 \\
 4232.44102(10) & 16547 & $-$0.00359 & $-$0.00078 & 9345.593549(8) & 244012 & $-$0.003817 & 0.000870 \\
 4233.42984(10) & 16591 & $-$0.00385 & $-$0.00103 & 9347.864040(7) & 244113 & $-$0.003691 & 0.001028 \\
 4236.55494(37) & 16730 & $-$0.00330 & $-$0.00050 & 9348.875513(7) & 244158 & $-$0.003767 & 0.000966 \\
 4237.45382(31) & 16770 & $-$0.00358 & $-$0.00079 & 9349.864424(7) & 244202 & $-$0.003925 & 0.000822 \\
 4238.44329(14) & 16814 & $-$0.00318 & $-$0.00039 & 9350.875784(8) & 244247 & $-$0.004113 & 0.000648 \\
 4246.37954(18) & 17167 & $-$0.00196 & 0.00080 & 9351.864576(8) & 244291 & $-$0.004391 & 0.000383 \\
 4247.36881(13) & 17211 & $-$0.00177 & 0.00099 & 9352.875857(8) & 244336 & $-$0.004658 & 0.000130 \\
 4254.38058(11) & 17523 & $-$0.00339 & $-$0.00067 & 9353.887113(8) & 244381 & $-$0.004951 & $-$0.000149 \\
 4255.34734(24) & 17566 & $-$0.00323 & $-$0.00051 & 9354.875881(7) & 244425 & $-$0.005253 & $-$0.000437 \\
 4265.32916(16) & 18010 & $-$0.00202 & 0.00065 & 9355.887182(7) & 244470 & $-$0.005500 & $-$0.000670 \\
 4266.27310(27) & 18052 & $-$0.00219 & 0.00048 & 9356.898528(7) & 244515 & $-$0.005703 & $-$0.000859 \\
 4267.26306(28) & 18096 & $-$0.00130 & 0.00137 & 9357.887463(8) & 244559 & $-$0.005837 & $-$0.000979 \\
 4268.34172(14) & 18144 & $-$0.00163 & 0.00104 & 9358.898958(8) & 244604 & $-$0.005890 & $-$0.001018 \\
 4269.24068(35) & 18184 & $-$0.00182 & 0.00084 & 9359.888029(7) & 244648 & $-$0.005889 & $-$0.001003 \\
\noalign{\smallskip}\hline 
\end{longtable}
\end{appendix}
\end{document}